# Probing Temperature at Nanoscale through Thermal Vibration Characterization using Scanning Precession Electron Diffraction

Kun Yang[1,#], Chao Zhang[1,#], Chengwei Wu[2,#], Qian Du[1], Bingzhi Li[3], Zhen Fang[1], Liang Li[4], Jianbo Wu[1], Tianru Wu[5,*], Hui Wang[2,*], Tao Deng[1,*], Wenpei Gao[1,*]

1 State Key Laboratory of Metal Matrix Composites, School of Materials Science and Engineering, Future Material Innovation Center, Zhangjiang Institute for Advanced Study, Shanghai Jiao Tong University, Shanghai 200240, China
2 School of Physics and Electronics, Hunan Key Laboratory of Super Microstructure and Ultrafast Process, State Key Laboratory of Powder Metallurgy, Central South University, Changsha 410083, China
3 Research Center for Scientific Data Hub, Zhejiang Lab, Hangzhou 310000, China
4 School of Physics and Astronomy, Shanghai Jiao Tong University, Key Laboratory for Particle Astrophysics and Cosmology (MOE), SKLPPC, Shanghai 200240, China
5 School of Mechanical Engineering, Shanghai Jiao Tong University, Shanghai 200240, China

**ABSTRACT.**
Accurate, non-contact temperature measurement with high spatial resolution is essential for understanding thermal behavior in integrated nanoscale devices and heterogeneous interfaces. However, existing techniques are often limited by the need for physical contact or insufficient spatial resolution for the measurement of local temperature and mapping its distribution. Here, we showcase the direct temperature measurement of graphene with nanometer spatial resolution in transmission electron microscopy. In experiments, combining a scanning nanobeam with precession electron diffraction offers the collection of kinematic diffraction from a local area at the nanometer scale. In analysis, we use a pre-calculated, sample-specific structure-factor-based correction method to enable the linear fitting of the diffraction intensities, allowing the determination of the Debye-Waller factor as a function of temperature at the precision of $10^{-4}$ Å$^2$/°C. With the high spatial resolution and measurement precision, the temperature and thermal vibration mapping further reveal the influence of graphene lattice parameters and thickness on the Debye-Waller factor, providing valuable insights into the vibrational properties impacted by temperature, lattice structure, and graphene layer thickness.

Thermal management in microelectronics has become a critical bottleneck as semiconductor devices continue to shrink in size and grow in complexity [1]. Mapping local temperature at characteristic dimension and at the atomic interfaces is essensial for evaluating the device performance. However, accurately measuring local temperature and mapping its distribution remains a challenge. Temperature measurement often involves indirect quantification through related physical quantities [2]. Traditionally, thermometric methods can be categorized into contact and non-contact types. Thermocouples fabricated into fine tips and integrated to atomic force microscopy can probe the temperature of sample surface [3], but requires physical contact and may introduce distortions. The spatial resolution is also limited by the physical size of the probe. Non-contact optical thermography techniques, including Raman spectroscopy [4], infrared spectroscopy [5], and fluorescence [6], do not require physical contact but are constrained by the diffraction limit [7], resulting in their spatial resolution of about 100s of nm to $\mu$m, larger than the size of state-of-the-art semiconductor functional units [8].

To overcome the limits, transmission electron microscopy (TEM) offers high spatial resolution in materials characterization. When coupled with established thermometric approaches, TEM can probe materials temperature by quantifying the nanoscale volumetric change of liquid metal confined in carbon nanotubes [9], or by characterizing the lattice thermal expansion of metallic nanoparticles [10]. Based on atomic resolution images, Hwang reported the quantification of local temperature in oxide by the attenuation or enhancement of atomic column intensity in HAADF images [11,12]. Additionally, in TEM, temperature can also be quantified using localized plasmon or phonon excitation in electron energy loss spectroscopy [13,14]. However, most available techniques in TEM can only be applied to samples of a specific type due to the detection or quantification limit of spectroscopy and imaging methods. These methods are either indirect measurement method, or still limited by resolution because a finite area of the sample needs to be included to acquire diffraction. A direct thermometric method with high spatial resolution is still missing in nanoscale metrology. Compared to the imaging methods, electron diffraction is simple to perform and less likely to be limited by the sample

In this letter, we report the measurement and mapping of local temperature using four-dimensional Scanning Transmission Electron Microscopy (4D-STEM) with a precessing beam. This method collects precession electron diffraction (PED) patterns during the probe scanning. By applying sample- and thickness-specific correction factors, a linear fit using the intensity computes the Debye-Waller factor directly, enabling temperature mapping and the probing of local vibration characteristics. Results from graphene show the capability of this method that can measure the Debye-Waller factor and therefore temperature at near 1 nm spatial resolution. Using lattice parameters and

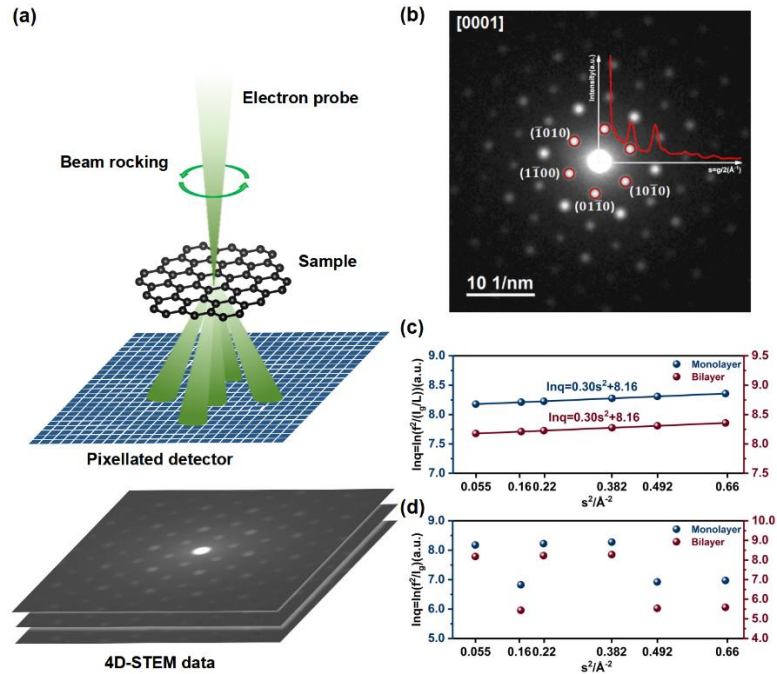

**Figure 1** (a) The experimental setup for 4D scanning electron diffraction using AC-STEM. (b) The average diffraction pattern from 4D electron diffraction dataset and the intensity profile of the pattern. (c) The Wilson plot of monolayer-graphene and AB-stacking bilayer graphene using the correction factor, L, from Table I and the results of linear fitting. (d) The Wilson plot without the correction factor, L.

selection. The only disadvantage is that electron diffractions typically do not offer high spatial resolution due to the size of the electron beam. With recent advances in aberration correction, an electron probe with the size as small as below 1 Ångstrom (Å) can be formed by converging the electron beam, offering atomic scale imaging resolution. Physical properties that can be probed with electron scattering should also be mappable at such resolution [15]. Therefore, we propose to probe the sample temperature with both high spatial resolution and improved precision using diffraction with a nanometer sized electron probe.

thickness determined by 4D-STEM, and combined with theoretical calculations, the work also reveals detailed insight into the vibrational properties governed by temperature, lattice, and graphene layer thickness.

Figure 1(a) shows the experimental set up of 4D-STEM with PED. Unlike conventional electron diffraction or imaging, during the raster scanning of the electron probe, the electron beam precesses about the optical axis at a constant angle at each probe position, forming a PED in the back focal plane. Each diffraction pattern is collected using a hybrid pixel electron detector oprated at the acquisition rate of 1000 frames per second (fps).

In electron diffraction analysis, a major obstacle is the dynamical effects that originates from electrons being scattered multiple times by the sample. Intensities in the dynamic diffraction patterns do not simply follow the kinematical theory, making property analysis based on the distribution of diffraction intensity challenging. Here, the net effect of PED is equivalent to the sample being precessed relative to a stationary axis. By averaging many reflections that satisfy the Bragg's law [16], ultimately, the effect of dynamic diffraction can be reduced and the diffraction intensity follows that of the kinematic diffraction. In kinematical theory, the $g_{th}$ order diffraction intensity, $I_g$, is proportional to the magnitude of the structure factor, $F(hkl)$, expressed as $I_g \propto \langle |F(hkl)|^2 \rangle$. The structure factor is also related to Debye-Waller factor, which is usually described as $e^{-Bs^2}$. Often B is simply called the Debye-Waller factor. We can express the diffraction intensity, $I_g$, as

$$KI_g = e^{-2Bs^2} f^2(s) \left[ \sum_{j=1}^{n} e^{2\pi i(hx_j + ky_j + lz_j)} \right]^2, \quad (1)$$

Detailed information can be found in supplementary information [17].

The Debye-Waller factor reflects the temperature-dependent phonon population, making it directly related to temperature. Therefore, measuring the Debye-Waller factor provides a direct means of determining the sample's temperature. While the Debye-Waller factor can usually be measured using X-ray, neutron diffraction, available methods require appropriate absorption and extinction correction, and rigorous data fitting [23-25]. Both X-ray and neutron scattering techniques are also limited by a spatial resolution of 100s of nm to micrometers at best. On the other hand, an accurate ab initio calculation of the Debye-Waller factor requires accurate value of the elastic constant and quantification of anharmonic effects, both are difficult to obtain [26-29].

To directly estimate Debye-Waller factor from diffraction, S. H. Yü [30,31] and A. J. C. Wilson [32] proposed that if we assume a random distribution of atoms in a unit cell, the log of intensity, $I_g$, devided by the summation of atomic scattering factors, $f^2(s)$, is propotional to the squre of the scattering vector, $s^2$, ie. $lnq = 2Bs^2 + lnK$, where $q = f^2(s)/(I_g/L)$, $K$ is a scaling factor, and L= $n^2$, where n is the number of atoms in the unit cell. Based on this, in the Wilson plot that reflects the X-ray diffraction intensity as a function of the scattering vector, $lnq$ and $s^2$ can be plotted as a straight line with a slope of 2B. Midgley et al. [33] extended this method in electron diffraction to measure the Debye-Waller factor from Higher-Order Laue Zone (HOLZ) reflections obtained by PED.

Using this method, however, even for crystals with a single element, the Wilson plot does not strictly satisfy a linear fit, because the the assumption of random distribution of atoms in the unit cell ignores the phase information in the diffraction caused by the positions of the atoms, resulting in the error in the value $L$. The descrepency leads to measurement errors and insufficient precision that cannot be amended or improved by experiments. Here we show because the $L$ from structure factor can be rigorously calculated, the effect of $L$ can be compensated using a correction factor $1/L$ that's determined only by the reciprocal lattice index and the sample thickness. In this work, we use monolayer graphene and AB-stacking bilayer graphene as model materials and measure the local temperature using Debye-Waller factor calculated from the corrected Wilson plot. We first calculate the correct $L$, as $L = [\sum_{j=1}^{n} e^{2\pi i(hx_j + ky_j + lz_j)}]^2$. The value for monolayer and AB-stacking bilayer graphene is listed in Table I.

**Table I**. $L$ of the $g_{th}$ order diffraction for monolayer-graphene and AB-stacking bilayer graphene

| crystal | 1st | 2nd | 3rd | 4th | 5th | 6th |
|---|---|---|---|---|---|---|
| monolayer | 1 | 4 | 1 | 1 | 4 | 4 |
| AB-bilayer | 1 | 16 | 1 | 1 | 16 | 16 |

By ploting the value, $lnq = \ln[f^2(s)/(I_g/L)]$, with the correct $L$, against $s^2$, the Wilson plot shows a perfect straight line in Figure 1(c). For comparison, Figure 1(d) shows the Wilson plot without such correction, which does not afford for high quality linear fit. Based on the slope of the straight line in Figure 1(c), we can directly obtain the Debye-Waller factor from a diffraction pattern.

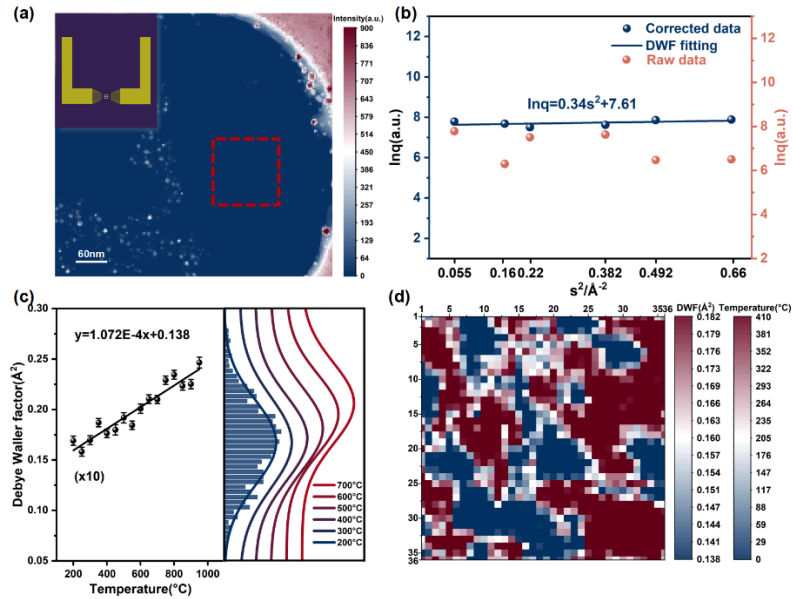

**Figure 2** (a) The annular dark field (ADF) image from 4D electron diffraction data by integrating the intensity of the annular region of 10-75mrad from the diffraction pattern Fig. 1(b). (b) The Wilson plot of the average diffraction intensities. (c) The Gaussian-fitting curve of different temperatures and the histogram plot of the red dashed area in (a) at a specified temperature (200 °C) and the Debye-Waller factor of monolayer graphene as a function of temperature. (d).The mapping of Debye-Waller factor of the red dashed area in (a) at 200 °C.

In experiment, monolayer and bilayer graphene samples are grown using a customized chemical vapor deposition method. For more detailed information about the procedures, please refer to the supplementary information [17]. The graphene sample was transferred to a MEMs based heating chip from Protochips Inc. for 4D-STEM with PED from room temperature to 950 °C. Using Thermo Scientific Spectra 300 aberration corrected STEM operated at 300 kV, the scanning PED was carried out by taking 4D-STEM using Topspin from NanoMegas under the micro probe diffraction mode with an aperture of 50 μm in diameter and different precession angles (0.02° for graphene, 2° for thicker graphite). The size of the scanning probe is measured to be about 1.39 nm. Each PED pattern is acquired with a Merlin hybrid pixel camera at the rate of 1000 frames per second. The top right inset of Figure 2(a) shows the PED pattern of a freestanding monolayer graphene. In a scanning PED experiment, we acquire diffraction patterns in 170 x 170 scanning positions with a step size of 3 nm, corresponding to a region of 510 nm x 510 nm in size, the calculated annular dark field image is shown in Figure 2(a). In the PED, the center position of the diffraction spots is first determined using circle edge fitting in Autodisk [21] and Centre of Mass (COM) method. For each diffraction spot, the radius is determined to be 4 pixels, within which the total intensity is counted as the diffraction intensity, details can be found in supplementary Figure S1. A region with clean, uniform image contrast is selected as indicated by the red dashed box in Figure 2(a) and the average diffraction pattern is shown in Figure 1(b). The average diffraction intensities are plotted as Wilson plot in Figure 2(b), similar to Figure 1(c,d), the corrected Wilson plot (in blue) fits better with a linear relationship than the uncorrected one (in red). Using the $L$ listed in Table I for monolayer graphene, a linear fit of the Wilson plot calculates the Debye-Waller factor from each diffraction pattern acquired at each location. At a specified temperature (e.g. 200 °C), the Debye-Waller factor obtained from 1600 diffraction patterns in the red dashed area were counted and plotted as a histogram, as shown in Figure 2(c), and fitted using Gaussian function. Figure 2(c) shows the Debye-Waller factor of the monolayer-graphene measured at different temperatures. The distribution of the measured Debye-Waller factor at different temperature is fitted with Gaussians and plotted in Figure 2(c). The center of Gaussians shift towards higher value with temperature.

The Debye Waller factors, when plotted against temperature, shows a clear linear relationship, with a slope of 1.072E-4 Å$^2$/°C. The mapping of Debye

Waller factor corresponding to the red dashed box in Figure 2(a) is shown in Figure 2(d) for 200 °C. Note that the 4D-STEM PED here offers a spatial resolution of about 1.39 nm, so temperature measurements can be obtained at the same resolution.

In Figure 2(d), it is seen that the Debye Waller factor fluctuates across the region. It is worth noting that freestanding graphene [34-36] may not be perfectly flat, ripples are often present in the surface. The humps and indents leads to a curved surface. In electron diffraction, the electron beam picks thermal vibration perpendicular to the incident direction. When the local graphene surface is tilted, the measured Debye-Waller factor reflects a summation of both the in-plane and out-of-plane thermal vibrations projected in the plane perpendicular to the electron beam. Therefore, the fluctuations in Debye-Waller factor measurements at the same temperature do not only originate from local temperature variations in graphene, but may also be related to its local surface contour.

We also estimated the effect of electron beam heating, which can be expressed as $\Delta T \approx I_0/e\kappa \cdot dE/dx$ [37-39], where $I_0$ is the beam current (15pA), $dE/dx$ is the energy loss rate per electron (0.15 eV/nm) and $\kappa$ is the thermal conductivity (5000 $W \cdot m^{-1} K^{-1}$ for graphene [40]). This value, 4.5E-7 K, is significantly smaller than the temperature measured based on diffraction intensity and hence can be ignored.

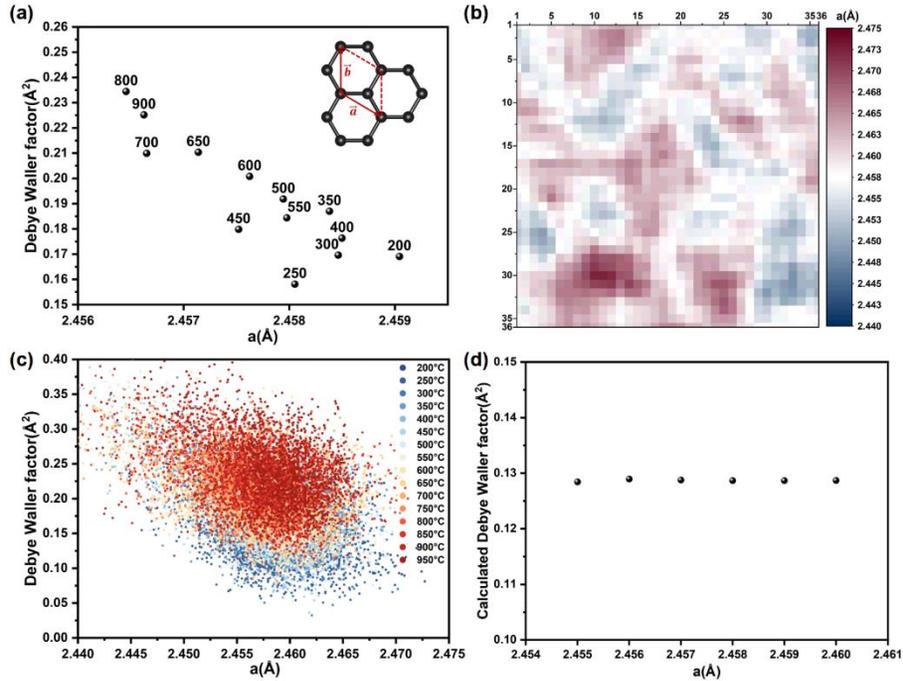

**Figure 3** (a) The Debye-Waller factor of monolayer graphene as a function of its lattice parameter, a. (b) The map of the lattice parameter mapping at 200 °C. (c) The scatter plot between DWF and lattice parameter at all temperatures. (d) The Debye-Waller factor as a function of monolayer-graphene's lattice parameter calculated by LAMMPS.

To study the thermal expension of graphene and how lattice parameter could impact Debye Waller factor [41-43], we first measure the lattice parameter from PED using AutoDisk [21]. From 200 to 950 °C, the lattice constant of graphene change from 2.459 to 2.456 Å as shown in supplementary Figure S2. In Figure 3(a), the lattice parameter of monolayer-graphene only changes about 1‰, while the Debye-Waller factor changes by about 33% from 0.169 to 0.225 Å$^2$. Figure 3(b) shows the mapping of the lattice at 200 °C in the same region as in Figure 2(d), no observable spatial correlation is seen in the two maps. At each temperature from 200 to 950 °C , the scatter plot of Debye-Waller factor and lattice parameter is presented in Figure 3(c), again, both the lattice parameter and Debye Waller factor change with temperature, but the two are not strongly correlated.

To better understand the impact of thermal expansion of graphene on Debye-Waller factor, we calculate the phonon density of states corresponding to different atomic models (with 0.2% strain applied within the graphene plane) using LAMMPS (supplementary Figure S3). Figure 3(d) shows the variation of calculated Debye-Waller factor with lattice parameter. The calculated Debye-Waller factor changes about 0.4%, when the lattice parameter change from 2.455 to 2.460 Å.

Both experimental and theoretical results show that the lattice change of monolayer graphene has almost a negligible effect on the measurement of Debye-Waller factor. The Debye Waller factor changes at a faster rate of about 0.044%/°C, 338 times higher than the change of lattice parameter (0.00013%/°C). In addition, lattice change is strongly contrained by the local atomic bonding, which may not be directly related to temperature. Since vibrational characteristics directly reflect temperature, temperature measurement using the Debye-Waller factor with 4D-STEM PED is more sensitive, reliable, and capable to offer nanometer-scale spatial resolution.

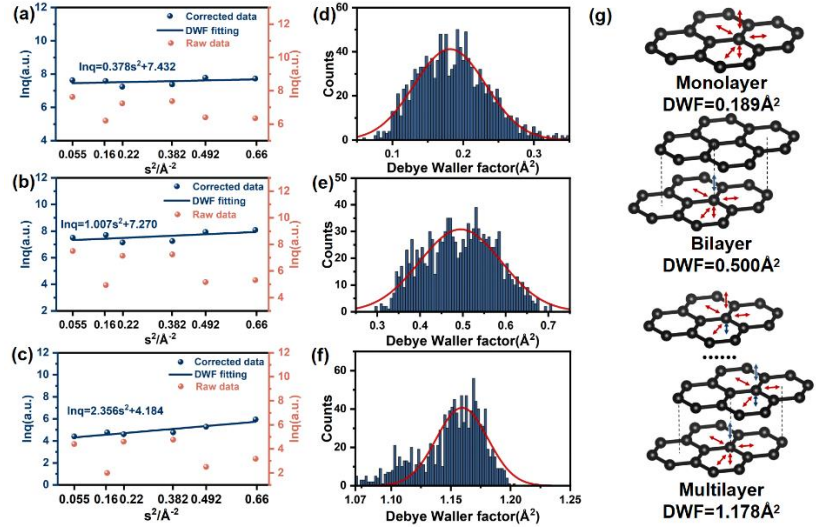

**Figure 4** Debye-Waller factor of different layer graphene at room temperature. The Wilson plot (a-c) and the histogram plot (d-f) of the selected area (see in Figure S4-area 1, area 2, Figure S5 ) of monolayer, bilayer and multilayer graphene. (g) The models of monolayer, bilayer, multilayer grapheneand the arrows indicate the direction of vibration of the carbon atoms.

In addition to lattice thermal expension, to investigate how the Debye-Waller factor changes with graphene thickness, we perform 4D-STEM PED on monolayer, bilayer and multilayer graphene at room temperature (Figure 4). The PED patterns of monolayer and bilayer graphene are also simulated using kinematical theory. The diffraction intensities were counted to compare with the experimental diffraction patterns as shown in Figure S4(b-c).

Multilayer graphene film is prepared by mechanically exfoliating thick graphite flakes. To identify the number of layers, position averaged convergent beam electron diffraction (PACBED) [44] is collected in experiment and compared with simulated patterns (supplementary Figure S5). The graphite is measured to be 11±1 nm in thickness. For AB-stacking multilayer graphene, the L of odd or even number of layers ($N_{layer}$=33±2 layers) is 15.95 or 16, respectively, we use 16 during the fitting [45]. For all three samples, monolayer, bilayer graphene, and thick graphite, as we expect, the Wilson plot fits well with a linear relationship after correction of the thickness specific L, while the uncorrected ones does not in Figure 4(a-c). Debye-Waller factors are measured using correction factors, L, from Table I, as shown in Figure 4(d-f).

The measured Debye Waller factors are 0.189, 0.500, and 1.178 Å² for monolayer, bilayer and multilayer graphene. Note that the Debye Waller factor of bilayer graphene is 2.645 times that of the monolayer, and graphite is 6.233 times.

To understand the change in measured Debye Waller factor in graphene with different thickness, we focus on the atomic bonding characteristics in layered materials. Depending upon the vibration direction, the vibration modes can be divided into the planar (in-plane) and the Z (out-of-plane) modes. In monolayer graphene, the carbon atoms have bonding only within the atomic plane, vibration along the out-of-plane direction is much stronger . Theoretical calculations of the mean square atomic displacements are $u_{xy}^2(0K)$ = 15.9 pm² and $u_z^2(0K)$ = 40.4 pm² [46]. In bilayer graphene, this out-of-plane vibration is much compressed and transferred to the in-plane direction, because of the coupling between the contacting surfaces. For a demonstration, in the Figure 4(g), the loss of bottom surface in the layer on the top, the loss of the top surface in the layer underneath, and the confinement applied against each surface reduced the Z mode vibration strength, whilst the same amount can be transferred in the planner mode. The total vibrational energy stays the same because it is solely determined by the kinetic energy as a function of temperature. This effect is more pronounced in graphite with higher thickness, as most of the Z-mode vibration is suppressed with the loss of surface, for the 33 graphene layers in this experiment, there are only two surfaces. Therefore, the Debye-Waller factor measurement of graphene is not only sensitive to temperature, but also reflect the vibration characteristics.

Here we have presented the temperature measurement with nanometer spatial resolution using 4D-STEM PED. With the corrected Wilson plot of the diffraction intensity, Debye-Waller factor of graphene can be calculated and is shown to be sensitive to temperature. The resolution of temperature mapping is determined and limited only by the size of the electron probe, which can be at the single nanometer level, significantly better than conventional macroscopic measurements such as Raman spectroscopy or thermocouples. The Debye Waller factor is also more sensitive to temperature than thermal expension. Both experiment and simulation further show that Debye-Waller factor reflects the vibration modes influenced by thickness and surface curvature, making it possible to probe the vibration characteristics using 4D-STEM PED.

## ACKNOWLEDGEMENTS

This work was supported by the National Key Research and Development Project of China (2024YFA1410600, 2021YFA1200801), the National Natural Science Foundation of China (grant nos. 52473237, 62174169), the Key Project of Frontier Science Research of Chinese Academy of Sciences (grant no. XDB30000000), and China Postdoctoral Science Foundation (2023M742225). The authors thank Prof. Jian-Min Zuo for his valuable suggestions and discussions.